\documentclass[aps,prl,twocolumn,groupedaddress,showpacs,floatfix,amsmath]{revtex4}
\usepackage{graphicx} 
\usepackage{latexsym} 
\usepackage{amssymb}  
\let\vec\boldsymbol    
\usepackage{psfrag}
\newcommand{\dbar}{{\mkern2.5mu\mathchar'26\mkern-12mu{d}\mkern1mu}}

\begin{document}


\title{Sensitivity of arrest in mode-coupling glasses to low-$q$ structure}


\author{M. J. Greenall, Th.\ Voigtmann, P. Monthoux and M. E. Cates}
\affiliation{SUPA, School of Physics, The University of Edinburgh, JCMB King's Buildings, Edinburgh EH9 3JZ, United Kingdom}


\date{\today}

\begin{abstract}
We quantify, within mode coupling theory, how changes in the liquid structure affect that of the glass. Apart from the known sensitivity to the structure factor $S(q)$ at wavevectors around the first sharp diffraction peak $q_0$, we find a strong (and inverted) response to structure at wavevectors \emph{below} this peak: an increase in $S(q_0/2)$ {\em lowers} the degree of arrest over a wide $q$-range. This strong sensitivity to `caged cage' packing effects, on length scales of order $2d$, is much weaker in attractive glasses where short-range bonding dominates the steric caging effect.
\end{abstract}


\maketitle

The mode-coupling theory (MCT) of the glass transition is a dynamical theory of viscous slowdown and arrest in glass-forming liquids \cite{leutheusser,goetze_rev}. It gives a good description of density relaxation in molecular and (especially) colloidal glass-formers \cite{goetze_exp}. However, the mathematical structure of MCT does not readily suggest a description in terms of the motion of individual particles, and does not lend itself easily to connections to other approaches, e.g.\ free energy landscapes (see e.g.\;\cite{stillinger}). Only the cage effect is recognized as a major ingredient to MCT \cite{weeks,goetze_rev}: glassy arrest within the theory is often driven by structural features on the nearest-neighbor lengthscale -- the structure factor peak value $S(q_0)$.

One major challenge is to understand how the structure of the glass-forming liquid affects the arrested state. We address this issue here, focussing on the role of density modulations at lengthscales beyond the near-neighbor spacing that defines the primary `cage'. Such modulations were seen in recent experiments and simulations on glass-formers that include silica melts, aromatic compounds, and gelation models \cite{comez,corezzi,horbach_prl, horbach_jpcm, puertas, fabbian}; their effect on the glass is poorly understood. 

Our approach mirrors one used in superconductivity theory (following \cite{monthoux}), in which, rather than compute the critical temperature $T_c$ for a given phonon spectrum, one sees how $T_c$ responds to small changes in this spectrum \cite{monthoux2}.  Arrest in glasses is characterized not only by the temperature at which ergodicity is lost, but by a wavelength-dependent nonergodicity parameter $F(q)$ (defined below). This describes the degree of arrest at different length scales, and gives the contribution to elastic scattering of the glass. We therefore choose to study the response of $F(q)$ to small changes in the static structure factor $S(q')$, since $F(q)$ is well-described by MCT even when there is significant error in the location of the glass transition in the phase diagram \cite{goetze_exp}.
\if{
(the analog of the phonon spectrum in superconductivity).
}\fi

Applying this method to repulsion-dominated glasses, we find an unexpected sensitivity to structural features at wavevectors {\em below} that of the main peak in $S(q)$. This is suggestive of a `caged cage' mechanism, in which structure just beyond the nearest neighbor shell is implicated in arrest. Modulation on this length scale, creating a positive feature in $S$ below the peak, leads to reduced freezing (smaller $F$) over a wide $q$-range: such modulation thus has a cage-softening effect. The same effect is almost absent in glasses stabilized by short-range attractions, reinforcing the idea that a distinct arrest mechanism (bonding not caging) is at work in such glasses. Hence, although in MCT $S(q)$ is the sole input to the glass transition, studying $\delta F(q)/\delta S(q')$ can reveal mechanistic subtleties that might otherwise be hard to discern. 

To begin, recall that MCT describes the relaxation of the Fourier modes (at wavevector $q$) of the time autocorrelations of the density fluctuations $\rho_q(t)$
\begin{equation}
\Phi_q(t)=\langle\rho_q(t)\rho^*_q(0)\rangle\,.
\label{density_corr}
\end{equation}
In the liquid, $F(q)=\lim_{t\rightarrow\infty}\Phi_q(t)$, is zero. MCT however predicts an idealized glass transition \cite{goetze_rev}. Here, the relaxation time of the density fluctuations diverges and they never decay completely, leading to a nonzero $F(q)$. The larger $F(q)$, the stronger the arrest at wavevector $q$.

The structure of the glass-forming liquid enters MCT through the static structure factor
$
S(q)=\langle\rho_q(0)\rho^*_q(0)\rangle
$
which, apart from the mean number density $n$, is the only input needed.
$F(q)$ satisfies \cite{goetze_rev}
\begin{equation}
 F(q) = S(q)M[F] (S(q)-F(q))\,,
\label{NEPimplicit}
\end{equation}
which follows from the $t\to\infty$ limit of the MCT equations of motion for correlators. Using a standard approximation for static three-point correlations, and factoring four-point dynamic ones, we have \cite{goetze_rev}
\begin{subequations}
\label{mctm}
\begin{align}
  M[F] &= \frac{n}{2q^2}\int\dbar^3k\,V(\vec q,\vec k)^2 F(k)F(p)
\\
  V(\vec q,\vec k) &= (\vec e_{\vec q}\cdot\vec k)c(k)+(\vec e_{\vec q}\cdot\vec p)c(p)\,,
\end{align}
\end{subequations}
where $\dbar k\equiv dk/(2\pi)$, $\vec e_{\vec q}=\vec q/|\vec q|$ is the
unit vector in the $\vec q$ direction, and $\vec p=\vec q-\vec k$. The direct correlation function
$c(q)$ obeys
$
  nc(q) = \left[1-S(q)^{-1}\right].
$

MCT gives a dynamical equation for $\Phi_q(t)$, in which $M[\Phi_q(t)]$ is a memory kernel that produces a frictional feedback mechanism, thought to be directly associated with cage formation \cite{goetze_rev}. This becomes stronger, eventually leading to the glass transition, as static correlations (governed by $c(q)$) increase.

We now consider only parameter values where $F(q)$ and $S(q)$ change smoothly with $n$ and $c(q)$. (This rules out the glass transition line itself; we return to this elsewhere \cite{TV}.)  Let us calculate $\delta F(q)/\delta S(q')$. This tells us how the degree of arrest at $q$ responds to a small change in $S(q')$ at $q'$. Equations (\ref{mctm}) can be written in terms of $q$, $k$ and $p$ \cite{franosch}. Considering small changes $\delta F(q)$ and $\delta S(q')$, we obtain an equation of the form
\begin{equation}
  \int \mathrm{d}k\,A(q,k)\delta F(k)
  =\int \mathrm{d}k\,B(q,k)\delta S(k)\,,
\label{deltaAB}
\end{equation}
where $A$ and $B$ are found from (\ref{NEPimplicit}) and (\ref{mctm}). The calculation of $\delta F(q)/\delta S(q')$ now reduces to the numerical inversion of $A(q,k)$.

All quantities $G(q,k)$ in (\ref{deltaAB}) separate into singular and regular contributions as:
\begin{equation}
G(q,k)=G(q)_\mathrm{sing}\delta(q-k)+G(q,k)_\mathrm{reg}\,.
\end{equation}
The singular contribution $(\delta F(q)/\delta S(q'))_\mathrm{sing}=F(q)(2S(q)-F(q))/S(q)^2\delta(q-q')$ gives the same-mode response to a change in $S(q)$. It ensures that the broad features of a change in $S(q)$, such as a shift in peak positions due to compression, will reappear in $F(q)$, although damped at higher $q$.

\begin{figure}
\psfrag{qpd}[][]{$q'd/2\pi$}
\psfrag{qd}[][][1][90]{$qd/2\pi$}
\psfrag{dF}[][][1][90]{$\delta F(q)/\delta S(q')$}
\psfrag{Sqp}[][][1][90]{$S(q')$}
\psfrag{0}[][]{$0$}
\psfrag{1}[][]{$1$}
\psfrag{2}[][]{$2$}
\psfrag{3}[][]{$3$}
\psfrag{4}[][]{$4$}
\psfrag{5}[][]{$5$}
\psfrag{6}[][]{$6$}
\includegraphics[width=.83\linewidth]{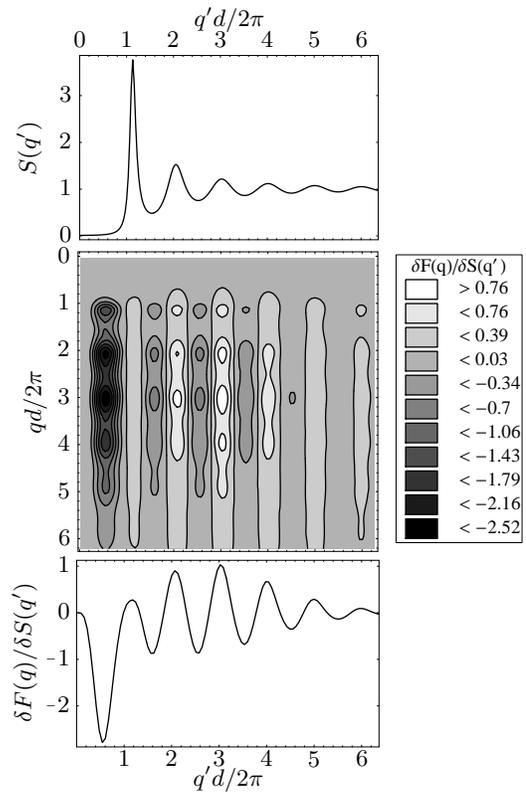}
\caption{\label{qfig}
Susceptibility of the hard-sphere-glass structure $F(q)$ to changes in the liquid structure factor $S(q')$. The middle panel shows the non-singular response $(\delta F(q)/\delta S(q'))_\mathrm{reg}$ (see text); the bottom panel a cut through $qd=19$ ($d$: diameter of the spheres). Top panel: $S(q')$ used in the calculation (Percus-Yevick approximation at volume fraction $\varphi\approx0.5236$).}
\end{figure}

\begin{figure}
\psfrag{qd}[][][1][90]{$qd/2\pi$}
\psfrag{rd}[][]{$r/d$}
\psfrag{gr}[][][1][90]{$g(r)$}
\psfrag{0}[][]{$0$}
\psfrag{1}[][]{$1$}
\psfrag{2}[][]{$2$}
\psfrag{3}[][]{$3$}
\psfrag{4}[][]{$4$}
\psfrag{5}[][]{$5$}
\psfrag{6}[][]{$6$}
\includegraphics[width=.78\linewidth]{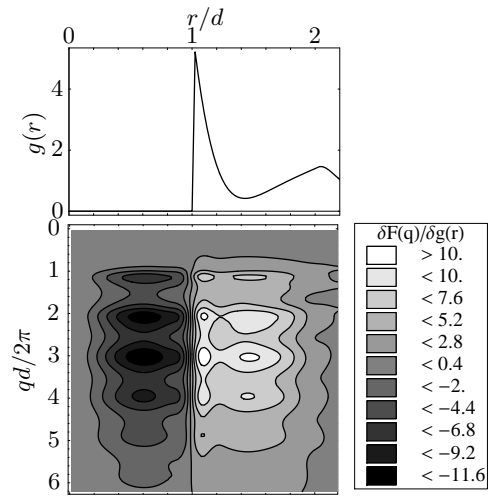}
\caption{Real space result $(\delta F(q)/\delta g(r))_\mathrm{reg}$ corresponding to Fig.\;\ref{qfig}. Upper panel: the radial distribution function $g(r)$ used in the calculation.}
\label{rfig}
\end{figure}

In contrast, the regular part of $\delta F(q)/\delta S(q')$ directly embodies the mode coupling concept -- a change in $S(q')$ at $q'$ affects $F(q)$ at all $q$. In Fig.\;\ref{qfig} we show $(\delta F(q)/\delta S(q'))_\mathrm{reg}$ for a system of hard spheres just within the glass. Fig.\;\ref{rfig} shows the corresponding $(\delta F(q)/\delta g(r))_\mathrm{reg}$, which gives the response of $F(q)$ to changes in the real-space radial distribution function
$
 g(r)=\int q\sin(qr)(S(q)-1)\,\mathrm{d}q/(4\pi^2 r n).
$

The dark patches of Fig.\;\ref{qfig} correspond to $(\delta F(q)/\delta S(q'))_\mathrm{reg}<0$, where a density modulation $\delta S(q')>0$ will reduce $F(q)$, softening the glass. In light areas, $(\delta F(q)/\delta S(q'))_\mathrm{reg}>0$, giving the opposite effect. The oscillations of the functional derivative are closely aligned with those of $F(q)$ and $S(q')$, with period near to the hard-core wavenumber $q_0=2\pi/d$ (cf.\ top panel and the constant-$q$ cut in the bottom panel of Fig.\;\ref{qfig}). We see that $\delta S(q')>0$ at $q'd/2\pi=1$ will increase $F(q)$ for (almost) all $q$. This confirms that strengthening the first peak in $S(q')$ promotes arrest. Less known is that this also holds for $q'd/2\pi=N$ with $N$ integer.

On top of the oscillations, there are two remarkable features of $\delta F(q)/\delta S(q')$.  First, Fig.\;\ref{qfig} shows a broad modulation envelope, maximal at relatively high $qd/2\pi\approx q'd/2\pi\approx 3$. The underlying period $q_0=2\pi/d$ means that this feature becomes localised ar $r\approx d$ in the real-space result (Fig.\;\ref{rfig}): it signifies strong sensitivity to changes in structure within a region of width $\approx 0.2d$ just beyond the hard-sphere diameter. Its $q$-space width scales linearly with $n$; we associate it with the localization length of the system. Since $\delta F(q)/\delta g(r)>0$ here, short-range bonding, which increases $g$, also increases $F$. \if{This might appear paradoxical, as such bonding can melt the glass, leading to a sharp jump of $F\to 0$ \cite{dawson,bergenholtz,pham}. However, assuming smooth variations with $c(q)$ keeps us in the glass, where $F$ is indeed increased by bonding (see e.g. \cite{dawson}).}\fi

The second striking feature of $\delta F(q)/\delta S(q')$ is a deep, long trough at $q'\approx q_0/2=\pi/d$ (Fig.\;\ref{qfig}). This tells us that adding $\delta S(q')>0$ at low $q'$ will reduce arrest over a wide range of $q$. This feature gives a modulation with period $\approx 2d$ in the real-space derivative (Fig.\;\ref{rfig}), helping to create an emphatic sign change at $r\simeq d$, such that that $\mathrm{sgn }(\delta F(q)/\delta g(r)) = \pm 1$ for $r \gtrless d$. This has two consequences: first, raising $g(r)$ for $r<d$, e.g.\ by softening the core of the potential, reduces $F(q)$ and weakens the glass, as expected. Second, the strongly positive region (at $r>d$) shows sensitivity to changes in $g(r)$ \emph{beyond} the nearest-neighbor shell. Fig.\;\ref{rfig} shows that the strongest sensitivity is at $r_\mathrm{min}\sim 1.5 d$, where $g(r)$ has a minimum; filling in this minimum strongly favors arrest.
Adding weight at $g(r_\mathrm{min})$ could be achieved by promoting `interstitial' particles between the first and second neighbor shells, strengthening the cages.
This `caged cage' effect emphasizes that the MCT glass transition is a collective phenomenon, and not just nearest-neighbor crowding, as many pictorial descriptions suggest. At the same time, the trough in Fig.\;\ref{qfig} shows that density modulations at $q'\sim\pi/d$, as might arise through incipient microphase separation \cite{puertas}, strongly soften the glass. (Such modulations might allow more local `rattle room' to develop.) 

\begin{figure}
\includegraphics[width=.78\linewidth]{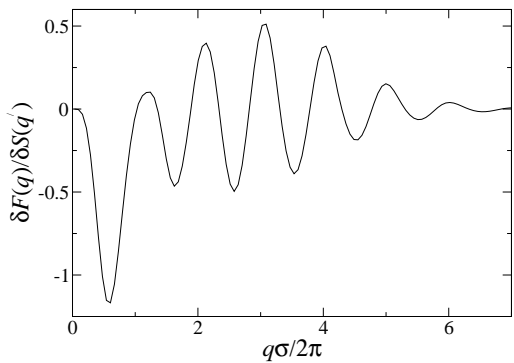}
\caption{Constant-$q$ cut through $(\delta F(q)/\delta S(q'))_\mathrm{reg}$ at the maximum amplitude for a Lennard-Jones system with PY closure. Temperature $T=\epsilon/k_\mathrm{B}$, with $\epsilon=1$ the well depth, and density $\rho=1.085/\sigma^3$, with $\sigma=1$ the molecular radius.}
\label{ljfig}
\end{figure}

These results are not specific to the hard-sphere system: to demonstrate this, we repeat the results of Fig.\;\ref{qfig} for a Lennard-Jones system of diameter $\sigma$ and attraction $\epsilon$ in Fig.\;\ref{ljfig} (only a fixed-$q$ cut is shown for simplicity). The same features -- a broad envelope peaked at $q\sigma/(2\pi)\approx3$ and a deep well at $q\sigma/(2\pi)\approx1$ -- are found even in this system where attractions typical of many molecular glass formers are present. A clear low-$q$ trough also remains in systems with rather stronger attraction, including hard-core Yukawa and broad \cite{dawson} square-well systems.

The occurrence of the trough can be rationalized by noting that the vertex $V$ in Eq.\;\ref{mctm} involves $S(q)$ mainly through $c(q)$. For $q\ll q_0$, $0<S(q)\ll1$ at high densities, as expected from the low compressibility of glass-forming liquids; thus, $\delta c/\delta S=1/(n S(q)^2)$ is large, and and one can expect $\delta V/\delta S=\delta V/\delta c\cdot\delta c/\delta S$ to be large as well. Note that $c(q)<0$, so $\text{sgn}(\delta V/\delta c)=-1$, and that $M[F]$ in Eq.\;\ref{NEPimplicit} is a positive functional, leading to the expectation $\delta F(q)/\delta S(q')\ll0$ for low $q'$.

\begin{figure}
\includegraphics[width=.78\linewidth]{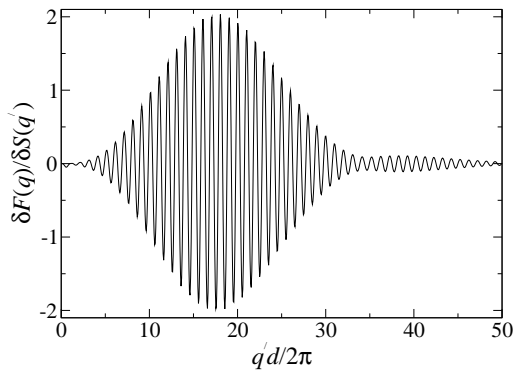}
\caption{Constant-$q$ cut through $(\delta F(q)/\delta S(q'))_\mathrm{reg}$ at the maximum peak height for a square well system at packing fraction $\varphi=0.5341$ and attraction depth $\epsilon=7.0k_\mathrm{B}T$. The width of the attraction $\delta=0.03$.}
\label{attfig}
\end{figure}

\begin{figure}
\includegraphics[width=.78\linewidth]{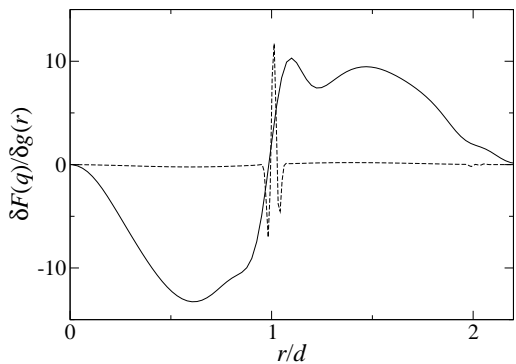}
\caption{Constant-$q$ cut through $(\delta F(q)/\delta g(r))_\mathrm{reg}$ at the maximum peak height for the square well system above (dashed line). Full line shows cut through hard-sphere system of Fig.\ \ref{rfig}.}
\label{attrfig}
\end{figure}

However, both the trough, and the consequences of it just outlined, are almost wiped out if strong and very short-ranged attractions are introduced into the system e.g.\ by supplementing the hard-sphere interaction with a narrow square well (Fig.\ \ref{attfig}).
As one enters the region of attraction-dominated glasses (see e.g.\ \cite{dawson,bergenholtz}), the trough shrinks in depth by a factor of around $100$. The broad envelope moves to very high $q$, corresponding to a very strong localization of the particles near to contact. Fig.\ \ref{attrfig} shows the qualitative change for a constant-$q$ cut through $(\delta F(q)/\delta g(r))_\text{reg}$. The hard-sphere results for $r\approx d$ (solid line) shows two main features: a peak of width $\approx0.2d$ corresponding to the $q$-space envelope (localization length), and broad peak of width $\approx d$ connected to the trough. In the square-well results (dashed line) the first peak has shrunk in width to $\sim\delta$, where $\delta$ is the attraction range. The second peak, describing the response to changes within the hard core, has almost vanished.  This reinforces the view that, in such glasses, arrest is primarily determined by bonding (enhancing the peak in $g(r)$ close to contact (Fig.\ \ref{attrfig})) rather than by the steric confinement effects seemingly responsible for the trough. We emphasize that the attraction needed to remove this trough is exceptionally strong and short-ranged, and could not (to the best of our knowledge) be realized in molecular systems.

Bearing this last point in mind, our study of $\delta F/\delta S$ may have several wider implications. Indeed there are many glass-forming systems showing a `pre-peak' or structural feature at $q<q_0$ \cite{comez,corezzi,horbach_prl,horbach_jpcm,puertas}.

Consider, for example, recent experimental and simulation work on sodium silicate melts just above their MCT glass transition \cite{horbach_prl,horbach_jpcm}. Here, the addition of sodium produces a pre-peak at $q<q_0$, and the glass is indeed weakened. This is driven by the formation of sodium-ion channels, but the general change of $F(q)$ with increasing pre-peak strength is nevertheless in line with our results. Also, adding a smaller species to a hard-sphere system increases $S(q)$ for low $q$ and in an appropriate parameter region produces a glass-softening effect \cite{williams,goetze_voigtmann}, again in line with our Fig.\;\ref{qfig}.

The effects of a pre-peak on $F(q)$ have also been studied in recent experiments on the glassformer $m$-toluidine \cite{comez}; these show the cusp predicted by MCT in the temperature dependence of $F(q)$ at various $q$. However, these X-ray diffraction studies do not reach high enough wavevectors to address the physics of the trough in $\delta F/\delta S$. Comparison might be possible with higher $q$ measurements, perhaps by neutron scattering.

A pre-peak also occurs in simulations of attractive colloids, where the potential was modified to suppress bulk phase separation and allow access to the glass transition \cite{puertas}. Reassuringly, our work suggests that in attraction-driven glasses a pre-peak does {\em not} strongly alter $F(q)$.

Low-$q$ structure, associated with arrested phase separation, may be seen in attractive glasses (high-density analogs of dilute colloidal gels) \cite{segre,shah,bergenholtz,ren, poonkroyreichman}. However, these features are at $q\ll q_0/2$ and our work is probably not relevant to them.
\if{
On another point, the form of $(\delta F(q)/\delta S(q'))_\mathrm{reg}$ also clearly demonstrates that the addition of polydispersity \cite{salgi} to a hard-sphere glass will lessen the degree of arrest. Here, the oscillations in $S(q')$ become weaker, and $S(q')$ increases at small $q$. The corresponding $\delta S(q')$ is completely `out of phase' with $(\delta F(q)/\delta S(q'))_\mathrm{reg}$, and $\delta F(q)$ will be largely negative.
}\fi

In conclusion, we have shown that the derivative $\delta F(q)/\delta S(q')$ is a useful probe of the sensitivity of the degree of arrest at one wavevector $q$ to changes in the glass-former's equilibrium structure at another wavevector $q'$. In repulsion-dominated glasses, there is strong response to structure in $S(q')$ at $q'$ below the first peak. Although formally we are dealing with infinitesimal perturbations, our results suggest that adding a `pre-peak' to $S(q)$ in this region may have a strong glass-softening effect (even independently of changes in the transition point itself). This effect, associated with packing effects beyond the nearest-neighbor shell, becomes much weaker in attraction-dominated glasses. Our predictions are non-trivial consequences of the MCT picture of the glass transition. Further experimental information on structural arrest in systems with low-$q$ modulations could therefore be valuable as a test of the overall applicability of the MCT picture.

MJG thanks A. Puertas for useful correspondence. This work was funded by EPSRC grant GR/S10377.


\begin{thebibliography}{99}
\bibitem{leutheusser} E. Leutheusser, Phys.\ Rev.\ A \textbf{29}, 2765 (1984).
\bibitem{goetze_rev} U. Bengtzelius, W. G{\"o}tze and A. Sj{\"o}lander, J.\ Phys.\ C \textbf{17}, 5915 (1984); W. G{\"o}tze, in \emph{Liquids, Freezing and the Glass Transition} (Les Houches Session LI), ed.\ J. P. Hansen, D. Levesque and J. Zinn-Justin (Amsterdam: North-Holland) (1991); W. G{\"o}tze and L. Sj{\"o}gren, Rep.\ Prog.\ Phys.\ \textbf{55}, 241 (1992).
\bibitem{goetze_exp} W. G{\"o}tze, J.\ Phys.\ Condens.\  Matter \textbf{11}, A1 (1999).
\bibitem{stillinger} P. G. Debenedetti and F. H. Stillinger, Nature \textbf{410}, 259 (2001).
\bibitem{weeks} E. R. Weeks and D. A. Weitz, Phys.\  Rev.\ Lett.\ \textbf{89}, 095704 (2002).
\bibitem{comez} L. Comez, S. Corezzi, G. Monaco, R. Verbeni and D. Fioretto, Phys.\ Rev.\ Lett.\ \textbf{94}, 155702 (2005).
\bibitem{corezzi} S. Corezzi, D. Fioretto and J. M. Kenny, Phys.\ Rev.\ Lett.\ \textbf{94}, 065702 (2005).
\bibitem{horbach_prl} A. Meyer, J. Horbach, W. Kob, F. Kargl and H. Schober, Phys.\ Rev.\ Lett.\ \textbf{93}, 027801 (2004).
\bibitem{horbach_jpcm} J. Horbach and W. Kob, J.\ Phys.\ Cond.\ Mat.\ \textbf{14}, 9237 (2002).
\bibitem{puertas} A. M. Puertas, M. Fuchs and M. E. Cates, J.\ Phys.\ Chem.\ B \textbf{109}, 6666 (2005); Phys.\ Rev.\ Lett.\  \textbf{88}, 098301 (2002).
\bibitem{fabbian} L. Fabbian {\em et al}., Phys.\ Rev.\ E \textbf{60}, 5768 (1999).
\bibitem{monthoux} G. Bergmann and D. Rainer, Z Phys. \textbf{263}, 59 (1973).
\bibitem{monthoux2} 
\if{
``Functional derivative techniques are found to help greatly in understanding the observed deviations from BCS laws'': 
}\fi
J. P. Carbotte, Rev.\ Mod.\ Phys.\ \textbf{62}, 1027 (1990).
\bibitem{TV} M. J. Greenall {\em et al}, in preparation
\bibitem{franosch} T. Franosch, M. Fuchs, W. G{\"o}tze, M. R. Mayr and A. P. Singh, Phys.\ Rev.\ E \textbf{55}, 7153 (1997).
\if{
\bibitem{lebowitz} J. L. Lebowitz and J. K. Percus, J.\ Math.\ Phys.\ \textbf{4}, 248 (1963), \emph{ibid} \textbf{4}, 116 (1963).
}\fi
\if{
\bibitem{girifalco} L. A. Girifalco, J.\ Phys.\ Chem.\ \textbf{96}, 858 (1992); J. M. Pacheco and J. P. Prates-Ramalho, Phys.\ Rev.\ Lett.\ \textbf{79},  3873 (1997).
}\fi
\bibitem{dawson} K. Dawson {\em et al.}, Phys.\ Rev.\ E \textbf{63}, 011401 (2001).
\bibitem{bergenholtz} J. Bergenholtz and M. Fuchs, Phys.\ Rev.\ E \textbf{59}, 5706 (1999).
\bibitem{pham} K. N. Pham {\em et al.}, Science\ \textbf{296}, 104 (2002).
\if{
\bibitem{barker} J. A. Barker and D. Henderson, J. Chem.\ Phys.\ \textbf{47}, 4714 (1967).
}\fi
\bibitem{williams} S. R. Williams and W. van Megen, Phys.\ Rev.\ E \textbf{64}, 041502 (2001).
\bibitem{goetze_voigtmann} W. G\"{o}tze and Th. Voigtmann, Phys.\ Rev.\ E \textbf{67}, 021502 (2003).
\bibitem{segre} P. N. Segr{\`e}, V. Prasad, A. B. Schofield and D. A. Weitz, Phys.\ Rev.\ Lett.\ \textbf{86}, 6042 (2001).
\bibitem{shah} S. A. Shah, S. Ramakrishnan, Y. L. Chen, K. S. Schweizer and C. F. Zukoski, Langmuir \textbf{19}, 5128 (2003).
\bibitem{ren} S. Z. Ren and C. M. Sorensen, Phys.\ Rev.\ Lett.\ \textbf{70}, 1727 (1993).
\bibitem{salgi} P. Salgi and R. Rajagopalan, Adv.\ Coll.\ Int.\ Sci.\ \textbf{43}, 169 (1993).
\bibitem{poonkroyreichman} K. Kroy, M. E. Cates and W. C. K. Poon, Phys.\ Rev.\ Lett.\ \textbf{92}, 148302 (2004); W. C. K. Poon {\em et al.}, Faraday Discuss. \textbf{ 112}, 143 (1999). 
\end{thebibliography}
\end{document}